\documentclass[conference]{IEEEtran}
\IEEEoverridecommandlockouts
\usepackage{cite}
\usepackage{amsmath,amssymb,amsfonts}
\usepackage{algorithmic}
\usepackage{graphicx}
\usepackage{textcomp}
\usepackage{xcolor}
\usepackage{xcolor}
\usepackage{geometry}
\graphicspath{{figures/}}

\newtheorem{theorem}{Theorem}
\newtheorem{corollary}{Corollary}
\newtheorem{definition}{Definition}
\newtheorem{lemma}{Lemma}
\newtheorem{assu}{Assumption}
\newtheorem{pb}{Problem}
\allowdisplaybreaks
\usepackage{algorithmic}
\def\BibTeX{{\rm B\kern-.05em{\sc i\kern-.025em b}\kern-.08em
    T\kern-.1667em\lower.7ex\hbox{E}\kern-.125emX}}
\begin{document}

\title{Sending Timely Status Updates through Channel with Random Delay via Online Learning
\\
\thanks{This work was supported by Tsinghua University-China Mobile Research Institute Joint Innovation Center. (\emph{Corresponding author: Jintao Wang})}
\thanks{The authors have provided public access to their code or data at https://github.com/loveisbasa/Infocom2022
}
}

\author{\IEEEauthorblockN{Haoyue Tang, Yuchao Chen, Jingzhou Sun, Jintao Wang, Jian Song}\\
\IEEEauthorblockA{\textit{Department of Electronic Engineering, Tsinghua University} \\
\textit{Beijing National Research Center for Information Science and Engineering, Beijing, China}\\
\textit{Research Institute of Tsinghua University in Shenzhen, Shenzhen, China}\\
\{thy17@mails, cyc20@mails, sunjz18@mails, wangjintao@mails, jsong@mails\}.tsinghua.edu.cn}
}

\maketitle

\begin{abstract}
In this work, we study a status update system with a source node sending timely information to the destination through a channel with random delay. We measure the timeliness of the information stored at the receiver via the Age of Information (AoI), the time elapsed since the freshest sample stored at the receiver is generated. The goal is to design a sampling strategy that minimizes the total cost of the expected time average AoI and sampling cost in the absence of transmission delay statistics. We reformulate the total cost minimization problem as the optimization of a renewal-reward process, and propose an online sampling strategy based on the Robbins-Monro algorithm. Denote $K$ to be the number of samples we have taken. We show that, when the transmission delay is bounded, the expected time average total cost obtained by the proposed online algorithm converges to the minimum cost when $K$ goes to infinity, and the optimality gap decays with rate  $\mathcal{O}\left(\ln K/K\right)$. Simulation results validate the performance of our proposed algorithm. 
\end{abstract}

\section{Introduction}
Timely status information is crucial for many real-time control system, e.g., the autonomous vehicular networks and the tactile internet. To measure the timeliness of status update, the metric Age of Information (AoI) \cite{roy_12_aoi} has been proposed. By definition, AoI measures the time elapsed since the freshest information stored at the receiver is generated, and a small AoI performance requires the system to possess both high throughput and low transmission delay \cite{roy_12_aoi}. However, this is often limited by the transmission resources of the sensor. Moreover, the lack of precise transmission statistics (e.g., channel condition and delay distribution) makes the design of status update system more challenging. 

Designing energy efficient sampling and transmission strategies to optimize the timeliness of status update systems have been studied in \cite{wang_21_twc,discrete_preempt,bo_sampling,tang_20_jsac,ceran_wcnc,aba_drl_aoi,arafa_online,bedewy_21_tit,sennur_gg1,najm,roy_15_isit,sun_17_tit}. In discrete time scenarios, minimizing the time average AoI performance can be formulated into a Markov decision process. When update packets are generated randomly by external environment, AoI minimum transmission and preemption strategies are proposed in \cite{wang_21_twc,discrete_preempt,bo_sampling,tang_20_jsac}. When the generation of update packets can be controlled at the transmitter, the joint design of sampling, power control and retransmission strategies are studied in \cite{ceran_wcnc,aba_drl_aoi,arafa_online}. In continuous time scenarios, when the status update generation is controlled by an external random process, the expected AoI performance under different service disciplines are analyzed in \cite{bedewy_21_tit,sennur_gg1}. When the update packets can be generated at will, low complexity algorithms to obtain the optimum sampling and transmission strategies are proposed in \cite{sun_17_tit,roy_15_isit}.

Notice that the aforementioned studies require the transmission statistics (e.g., transmission delay or packet-loss probabilities) to be known in advance. When such information is unavailable to the transmitter, reinforcement learning is an efficient tool to learn the optimum policy adaptively based on historical transmissions. Reinforcement learning algorithms such as Q-learning and SARSA have been employed to obtain the AoI minimum sampling and transmission strategies \cite{ceran_19_infocomwks,ceran_21_jsac}. To reduce the storage and computational complexity, SARSA with tile coding \cite{kam_rl}, deep Q-Learning \cite{aba_drl_aoi} and actor-critic algorithm \cite{aylin_rl} have been used to tackle with the huge state space of AoI minimization problems. The aforementioned RL based algorithms require the transmitter either to store a large table of value functions, or to train a neural network for function approximation. Either exerts high storage and computational burden to the sensor. Moreover, analyzing the convergence rate of the proposed algorithms remains challenging. 

Online learning and bandit algorithms provide efficient solutions with a low computational cost for sequential decision making in an unknown environment. 
When the generation of update packets are controlled by external environment and arrives randomly, AoI minimum adaptive channel selection and link scheduling algorithms based on bandit algorithms have been proposed \cite{aoibandit,atay2020aging,banerjee_adversarial_aoi}. Vishrant \emph{et al.} \cite{tripathi2021online} model the timeliness at the receiver as a time-varying function of the AoI, and propose online learning algorithms that can adapt to adversarial cost function variations. In multi-user networks, \cite{li2021efficient} proposes scheduling algorithms that can satisfy the timeliness constraint of each user with sublinear cumulative utility regret. However, the aforementioned studies deal with discrete time scenarios and assume the transmission of each packet is instantaneous, i.e., the transmission delay is one slot or can be ignored. Although an online sampling algorithm for AoI minimization over a two-way random delay has been proposed in \cite{chichun-19-isit}, the regret performance is not well understood. 

To solve this problem, we consider a point-to-point status update system with a sensor sampling and transmitting information updates to the receiver through a lossless network with a random transmission delay, similar to \cite{sun_17_tit,arafa_model}. We assume each transmission attempt incurs an extra cost, and aim at minimizing the total cost that consists of both the average AoI and transmission cost. The main
contributions of the paper are as follows:
\begin{itemize}
	\item By reformulating the average cost minimization problem as a renewal-reward process optimization, we derive the optimum off-line sampling strategy when the transmission delay distribution is known. 
	
	\item We propose an online adaptive sampling strategy when the transmission delay distribution is unknown using the Robbins-Monro algorithm \cite{robbins_monro,neely2021fast}. Denote $K$ to be the number of samples we have taken, we show that the gap between the cumulative expected cost by using the proposed online algorithm and the minimum cost decays like $\mathcal{O}\left(\ln K/K\right)$, i.e., the proposed online algorithm adaptively learns the optimal policy. 
\end{itemize}

\section{Problem Formulation}

\subsection{System Model}
\begin{figure}[h]
	\centering
	\includegraphics[width=.47\textwidth]{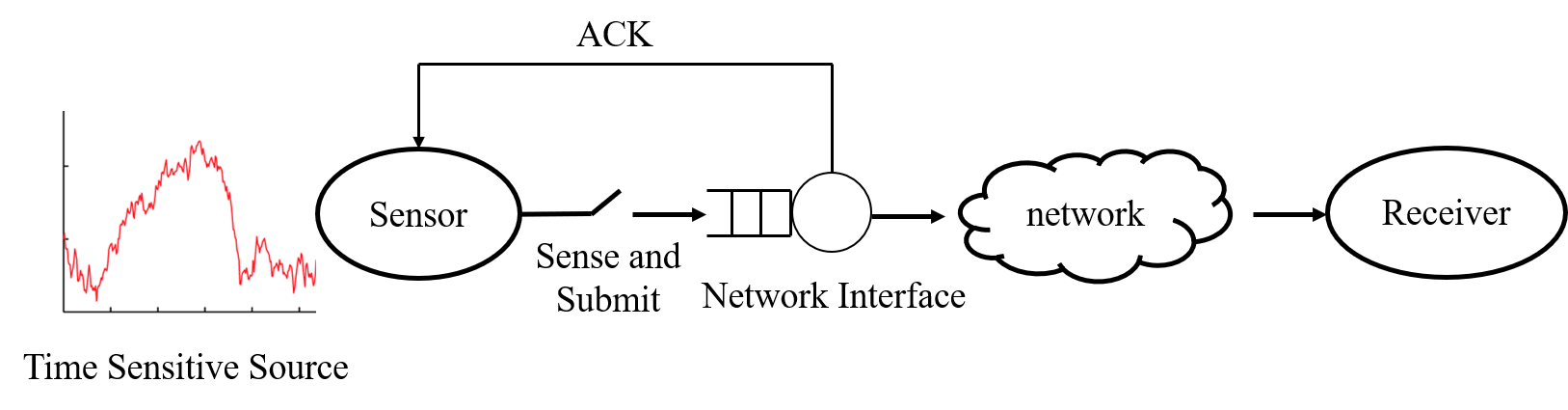}
	\caption{System model. }
	\label{fig:model}
\end{figure}

As is depicted in Fig.~\ref{fig:model}, we consider a sensor observes a time-sensitive process, samples and transmits update information to the receiver through a network interface queue similar to \cite{sun_17_tit,arafa_model}. The network interface serves the update packets on the First-Come-First-Serve (FCFS) basis. An ACK will be sent back to the sensor once an update packet is cleared at the interface, and we assume the transmission duration after passing the network interface is negligible. 

Suppose the sensor can generate and submit update packets at any time $t\in\mathbb{R}^+$ at will. Notice that update packets become stale while waiting in the queue, and the busy/idle information of the queue is available to the sensor through the ACK. Therefore, it is better to sample a new update packet after the ACK of the previous submitted packet is received. Denote $S_k$ to be the generation time-stamp of the $k$-th update packet and the time duration spent in the network interface is $D_k$. We assume each $D_k$ is independent and identically distributed (i.i.d.) following probability measure $\mathsf{P}_D$.
\begin{assu}
	The probability measure $\mathsf{P}_D$ is absolutely continuous. Moreover, its expectation and second order moment is bounded, i.e., 
	\begin{subequations}
	\begin{align}
		&0<\overline{D}_{\text{lb}}\leq\overline{D}\triangleq\mathbb{E}_{\mathsf{P}_D}[D]\leq\overline{D}_{\text{ub}}<\infty,\\ &0<M_{\text{lb}}\leq\mathbb{E}_{\mathsf{P}_D}[D^2]\leq M_{\text{ub}}<\infty,
	\end{align}
	\end{subequations}
	where $\overline{D}_{\text{lb}}, \overline{D}_{\text{ub}}$ are the lower and upper bound of the average transmission delay. The lower and upper bound of the second order moment of transmission delay are denoted by $M_{\text{lb}}, M_{\text{ub}}$, respectively.
\end{assu}

Since the queue is empty when the $k$-th update packet is submitted, packet $k$ will be received at time $S_k+D_k$. Since each update packet $k$ is generated after the reception of packet $k-1$, we denote $W_{k+1}:=S_{k+1}-(S_{k}+D_{k})$ as the ``\emph{waiting}'' time to take the $k+1$-th sample after receiving the ACK of the $k$-th sample at time $S_{k}+D_{k}$. 
\subsection{Age of Information}
We measure the information freshness of the receiver at time $t$ via the Age of Information \cite{roy_12_aoi}, namely the time elapsed since the latest information stored at the receiver is generated. 
Let $i(t):=\arg\max_{k\in\mathbb{N}}\{k|S_k+D_k\leq t\}$ be the index of the latest sample received by the destination before time $t$. 
The AoI at time $t$, denoted by $A(t)$ is:
\begin{equation}
	A(t):=t-S_{i(t)}. 
\end{equation}

\begin{figure}[h]
	\centering
	\includegraphics[width=.47\textwidth]{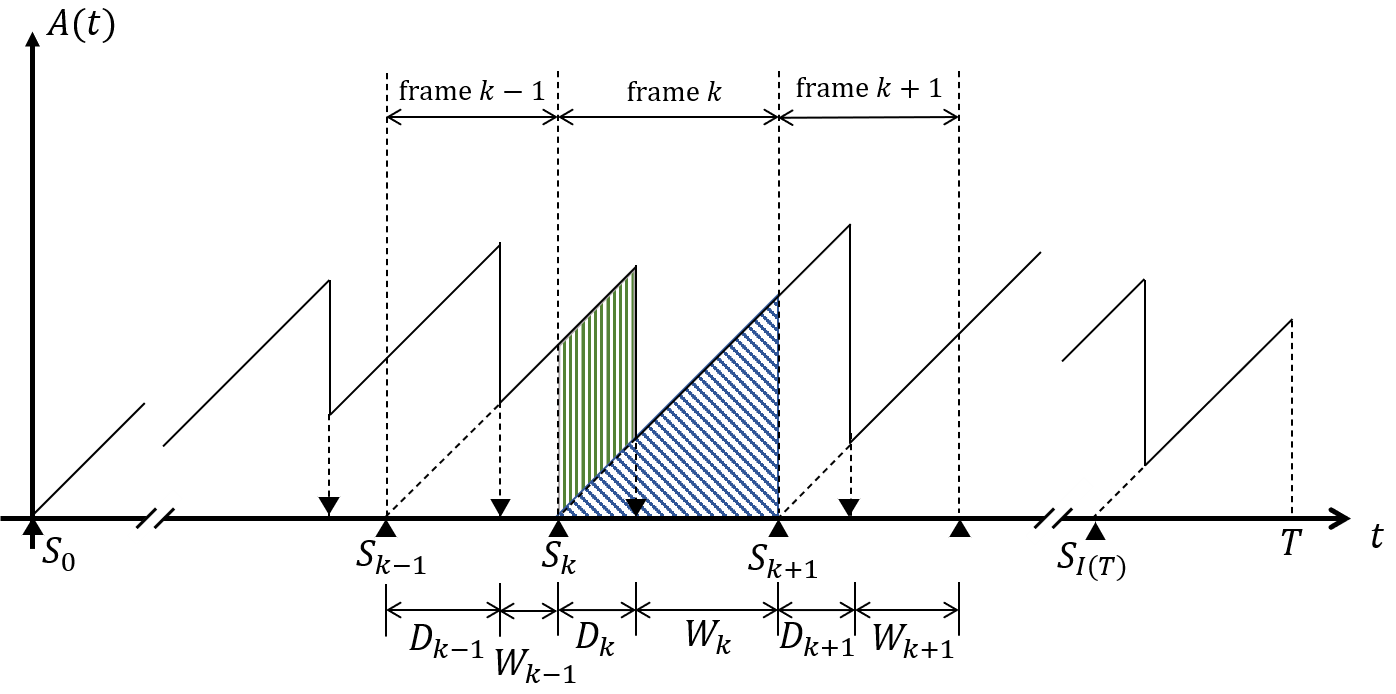}
	\caption{Illustration of AoI evolution.}
	\label{fig:aoievolve}
\end{figure}

A sample path of AoI evolution is depicted in Fig.~\ref{fig:aoievolve}. 

\subsection{Optimization Problem Formulation}
Denote $\overline{A}_{\pi,T}$ and $\overline{F}_{\pi, T}$ to be the expected average AoI and sampling frequency by using policy $\pi$ over an interval $(0, T)$, i.e., 
\[\overline{A}_{\pi, T}:=\mathbb{E}\left[\frac{1}{T}\int_{t=1}^TA(t)\mathsf{d}t\right],\overline{F}_{\pi, T}:=\mathbb{E}\left[\frac{i(T)}{T}\right].\] 

We assume an extra cost of $C\geq 0$ is incurred each time when the sensor samples and submits an update packet. Assume that $\mathsf{P}_D$ is unknown to the transmitter, the goal is to design a sampling strategy $\pi$ represented by a set of waiting times $\{W_k\}_{k=1}^{\infty}$ to minimize the total sum of expected average AoI and sampling cost based on $\overline{D}_{\text{lb}}, \overline{D}_{\text{ub}}, M_{\text{lb}}, M_{\text{ub}}$. Specifically, we focus on the class of causal policies denoted by $\Pi_{\text{Causal}}$, where each policy $\pi\in\Pi_{\text{Causal}}$ selects the waiting time $W_k$ of the $(k+1)$-th update packet based on the transmission delay of the $k$-th packet $D_k$ and the historical transmissions denoted by filtration $\mathcal{F}_{k-1}:=\sigma\left(\{(D_\kappa, W_{\kappa})\}_{\kappa=1}^{k-1}\right)$, where $\sigma(\cdot)$ represents the $\sigma$-field generated by random variables. 
The overall optimization problem is as follows:
\begin{pb}{\label{pb:rr-reformulate}}
\begin{subequations}
	\begin{align}
	&\pi^*\triangleq\mathop{\arg\min}\limits_{\pi\in\Pi_{\text{Causal}}}h_\pi:=\left(\overline{A}_\pi+C\overline{F}_\pi\right),\label{eq:primalobj}\\
	&\hspace{-0.2cm}\text{where }\overline{A}_\pi:=\mathop{\lim\sup}\limits_{T\rightarrow\infty}\overline{A}_{\pi,T}=\mathop{\lim\sup}\limits_{T\rightarrow\infty}\mathbb{E}\left[\frac{1}{T}\int_{t=1}^TA(t)\mathsf{d}t\right]\!,\\
	&\hspace{0.7cm}\overline{F}_\pi:=\mathop{\lim\sup}\limits_{T\rightarrow\infty}\overline{F}_{\pi, T}=\mathop{\lim\sup}\limits_{T\rightarrow\infty}\mathbb{E}\left[\frac{i(T)}{T}\right].
	\end{align}
\end{subequations}
\end{pb}

\section{Problem Resolution}
In this section, we first reformulate Problem 1 as the optimization of a renewal-reward process. In Subsection-B, we derive the optimum offline policy when the distribution $\mathsf{P}_D$ is known, which provides design insight and is used to measure the convergence rate of the proposed online algorithm. And then provide an adaptive online sampling strategy in Subsection-C when $\mathsf{P}_D$ is unknown. Finally, we analyze the average AoI performance of the proposed online policy in Subsection-D. 
\subsection{A Renewal-Reward Process Reformulation}
To formulate Problem 1 as a renewal-reward process optimization problem, we define ``\emph{frame}'' $k$ to be the interval of time during the generation of the $k$-th and the $(k+1)$-th update packet. The length of the $k$-th frame is:
\begin{equation}
L_k:=
D_k+W_k.
\end{equation}

According to Fig.~\ref{fig:aoievolve}, the cumulative AoI in frame $k$, denoted by  $X_k:=\int_{t=S_{k}}^{S_{k+1}}A(t)\mathsf{d}t$, is the sum of the area of a parallelogram and a triangle, i.e., 
\begin{align}
X_k=
(D_{k-1}+W_{k-1})D_k+\frac{1}{2}L_k^2.\label{eq:reward}
\end{align}

For ease of exposition, we denote $D_0=0$ and $W_0=0$. We consider
the time interval $(0, T)$ with $T=S_{K+1}$, i.e., the time when the $(K+1)$-th update packet is sampled. Let $Y_k=X_k+C$ be the sum of cumulative AoI and sampling cost in frame $k$, then with probability 1, the total cost in \eqref{eq:primalobj} can be rewritten as follows:
\begin{align}
h_\pi=&\lim_{K\rightarrow\infty}\frac{\mathbb{E}_\pi[\sum_{k=1}^KX_k]}{\mathbb{E}_\pi[\sum_{k=1}^KL_k]}+\lim_{K\rightarrow\infty}\frac{C\cdot K}{\mathbb{E}_\pi[\sum_{k=1}^KL_k]}\nonumber\\
\overset{}{=}&\lim_{K\rightarrow\infty}\frac{\mathbb{E}_\pi[\sum_{k=1}^KY_k]}{\mathbb{E}_\pi[\sum_{k=1}^KL_k]}.\label{eq:averageAoIreformulate}
\end{align}

\begin{definition}
	Let $\Pi_{\text{SD}}\subset\Pi_{\text{Causal}}$ be the set of stationary deterministic policies. A stationary deterministic policy $\pi\in\Pi_{\text{SD}}$ chooses the waiting time $W_k$ in frame $k$ based on a deterministic function parameterized by policy $\pi$ and the current delay, i.e.,  $W_k=f_\pi(D_{k}), \forall k$. 
\end{definition}
\begin{theorem}
	If the distribution $\mathsf{P}_D$ is known, there exists a stationary policy that is optimal to Problem~\ref{pb:rr-reformulate}. 
\end{theorem} 
\begin{IEEEproof}
	The proof is similar to \cite[Theorem 2 and Theorem 3]{sun_17_tit} and is omitted due to space limitations. 
\end{IEEEproof}

We then focus on searching for the optimum stationary policy $\pi\in\Pi_{\text{SD}}$. With slight abuse of notations, we denote $\pi(D)$ to be the waiting time selected by a policy $\pi$ upon observing transmission delay $D$. To facilitate further analysis, denote
\begin{equation}R_k:=\frac{1}{2}L_k^2+C.
\end{equation}
Therefore the total cost in frame $k$ can be written as \begin{equation}
	Y_k=R_k+L_{k-1}D_{k}\label{eq:tot-frame}
\end{equation} Then, the optimization objective \eqref{eq:averageAoIreformulate} can be rewritten as follows:
\begin{align}
	h_\pi=&\lim_{K\rightarrow\infty}\frac{\mathbb{E}[\sum_{k=1}^K(\frac{1}{2}L_{k}^2+L_{k-1}D_k+C)]}{\mathbb{E}[\sum_{k=1}^KL_k]}\nonumber\\
	=&\lim_{K\rightarrow\infty}\frac{\mathbb{E}[\sum_{k=1}^KR_k]+\mathbb{E}[\sum_{k=1}^KL_{k-1}D_k]}{\mathbb{E}[\sum_{k=1}^KL_k]}\nonumber\\
	\overset{(a)}{=}&\lim_{K\rightarrow\infty}\frac{\mathbb{E}[\sum_{k=1}^KR_k]+\mathbb{E}[\sum_{k=1}^KL_{k-1}]\overline{D}}{\mathbb{E}[\sum_{k=1}^KL_k]},\label{eq:objprev}
\end{align}
where $(a)$ holds because $D_k$ is i.i.d. 

We treat $R_k$ as the ``\emph{reward}'' in frame $k$. For stationary deterministic policy $\pi\in\Pi_{\text{SD}}$, the reward $R_k$ and length $L_k$ of frame $k$ only depend on the current delay $D_k$. Since the delay $D_k$ of each packet is i.i.d, the reward and frame length $(R_k, L_k)$ is independent of $(R_{k'}, L_{k'})$ in other frames and can be modeled as a renewal-reward Process. Moreover, the expectation $\mathbb{E}[L_k]$ is bounded. Therefore, according to the renewal theory, denote $\mathbb{E}[R]=\mathbb{E}[\frac{1}{2}(D+\pi(D))^2+C]$and $\mathbb{E}[L]=\mathbb{E}[D+\pi(D)]$, with probability 1, Problem~\ref{pb:rr-reformulate} can be reformulated as follows:

\begin{pb}[Renewal-Reward process reformulation of Problem 1]\label{pb:primal-rr}
	\begin{align}
	\pi^*
	=&\mathop{\arg\min}\limits_{\pi\in\Pi_{\text{SD}}}\left(\frac{\mathbb{E}[\frac{1}{2}(D+\pi(D))^2+C]}{\mathbb{E}[D+\pi(D)]}+\overline{D}\right). 
	\end{align}
\end{pb}

\subsection{Optimal Offline Algorithm based on $\mathsf{P}_D$}
We first analyze the optimum stationary policy when the delay distribution $\mathsf{P}_D$ is known. The optimum offline policy will
provide important insight to the design of online algorithm. 

Recall that $h_{\pi^*}$ is the minimum cost any policy $\pi\in\Pi_{\text{SD}}$ can achieve, i.e., \begin{equation}\frac{\mathbb{E}_\pi[\frac{1}{2}(D+\pi(D))^2+C]}{\mathbb{E}_\pi[D+\pi(D)]}+\overline{D}\geq h_{\pi^*}, \forall \pi\in\Pi_{\text{SD}}.\label{eq:renewalrewartopt}
\end{equation}

Denote $\gamma^*=h_{\pi^*}-\overline{D}$. Multiplying $\mathbb{E}_\pi[D+\pi(D)]$ on both sides of inequality \eqref{eq:renewalrewartopt}, we have:
\begin{equation} \frac{1}{2}\mathbb{E}[(D+\pi(D))^2+C]-\gamma^*\mathbb{E}[D+\pi(D)]\geq 0, \forall \pi\in\Pi_{\text{SD}}. \label{eq:opt-inequality}\end{equation}

For simplicity, denote function
\[g(\pi, \gamma):=\frac{1}{2}\mathbb{E}[(D+\pi(D))^2+C]-\gamma^*\mathbb{E}[D+\pi(D)].\]
When $\gamma=\gamma^*$, inequality \eqref{eq:opt-inequality} implies $g(\pi, \gamma^*)=0$ if and only if $\pi=\pi^*$. Therefore, if the $\gamma^*$ is known, the optimum sampling policy that achieves $h_{\pi^*}$ can be obtained by solving the following constrained optimization problem:
\begin{pb}[Functional Optimization Problem]\label{pb:c-rr-problem}
\begin{align}
	&\min_{\pi\in\Pi_{\text{SD}}} g(\pi, \gamma^*):=\nonumber\\
	&\hspace{0.5cm}\mathbb{E}\left[\frac{1}{2}(D+\pi(D))^2+C-\gamma^*(D+\pi(D))\right].\label{eq:consopt}
\end{align}
\end{pb}

To search for the optimum policy $\pi^*$, we present the following lemmas and corollaries:
\begin{theorem}\label{thm:threshold}
	Denote $\tilde{\pi}_{\gamma}^*:=\arg\min_{\pi\in\Pi_{\text{SD}}}g(\pi, \gamma)$ as the optimum stationary deterministic policy that minimizes the function $g(\pi, \gamma)$. Policy $\tilde{\pi}_{\gamma}^*$ specifies the waiting time upon observing transmission delay $D$ as follows:
		\[\tilde{\pi}_{ \gamma}^*(D)=\left(\gamma-D\right)^+.\]
\end{theorem}
\begin{IEEEproof}
The strict mathematic proof is similar to and is omitted due to space limitations \cite{sun_17_tit}.
\end{IEEEproof}

\begin{corollary}\label{coro:gamma-bound}
	The optimum ratio $\gamma^*=\frac{1}{2}\frac{\mathbb{E}\left[(D+\pi^*(D))^2\right]}{\mathbb{E}[D+\pi^*(D)]}$ and can be lower and  upper bounded by:
	\begin{align}
	\gamma_{\text{lb}}:=\frac{1}{2}\overline{D}_{\text{lb}}\leq\gamma\leq\frac{\frac{1}{2}M_{\text{ub}}+C}{\overline{D}_{\text{lb}}}=:\gamma_{\text{ub}}.
	\end{align}
\end{corollary}
\begin{corollary}\label{coro:uniquesolve}
	Define mapping $\mathcal{T}:\mathbb{R}\rightarrow\mathbb{R}$ to be:
	\begin{equation}
		\mathcal{T}(\gamma):=\frac{\mathbb{E}[\frac{1}{2}(D+\tilde{\pi}_{ \gamma}^*(D))^2+C]}{\mathbb{E}[D+\tilde{\pi}_{ \gamma}^*(D)]}.
	\end{equation}
	Let $\mathcal{T}^{(\tau)}(\gamma)\triangleq\underbrace{\mathcal{T}\circ\cdots\circ\mathcal{T}}_{\tau\text{ times}}(\gamma)$. Then $\lim_{\tau\rightarrow\infty}\mathcal{T}^{(\tau)}(\gamma)=\gamma^*, \forall \gamma\in[\gamma_{\text{lb}}, \gamma_{\text{ub}}]$. 
\end{corollary}

Proofs for Corollary~\ref{coro:gamma-bound} and \ref{coro:uniquesolve} are provided in Appendix~\ref{pf:coro:gamma-bound} and \ref{pf:coro:uniquesolve}, respectively. 

We then present our offline algorithm that find $\gamma^*$ iteratively using Corollary~\ref{coro:uniquesolve} if the delay distribution $\mathsf{P}_D$ is known:
\begin{itemize}
    \item We initialize $\gamma_0$ by uniformly choosing from interval $[\gamma_{\text{lb}}, \gamma_{\text{ub}}]$. 
    \item In each iteration $j\geq 1$, we compute the corresponding mapping $\gamma_j=\mathcal{T}(\gamma_{j-1})$. The iteration stops in iteration $J$ when the absolute value is below a certain threshold $|\gamma_J-\gamma_{J+1}|\leq \delta$.
    \item We use the policy $\tilde{\pi}_{\gamma_J}^*$ in the last epoch, i.e., when observing transmission delay $D$, we wait for $W=(\gamma_J-D)^+$ before taking the next sample. 
\end{itemize}

\subsection{Proposed Online Algorithm}
We then provide an online AoI minimization algorithm in the absence of distribution $\mathsf{P}_D$. The key is to maintain a sequence $\{\gamma_k\}$ that reflects our guess of the optimum $\gamma^*$ using the Robbins-Monro algorithm \cite{robbins_monro}. We initialize $\gamma_{0}\in[\gamma_{\text{lb}}, \gamma_{\text{ub}}]$ randomly in the first frame $k=1$. For frame $k\geq 2$, the algorithm operates as follows: 
\begin{subequations}
\begin{itemize}	

\item We observe the transmission delay $D_{k}$ and choose waiting time:
\begin{equation}
    W_{ k}=\left(\gamma_{k}-D_{ k}\right)^+. 
\end{equation}
We then compute the frame length $L_{k}=D_{k}+W_{k}$ and $R_{k}=\frac{1}{2}L_{k}^2+C$. 

\item  We then update $\gamma_{k}$ via the Robbins-Monro algorithm \cite{robbins_monro} as follows:
\begin{align}
&\gamma_{k+1}=\left[\gamma_{k}+\eta_k\left(R_{k}-\gamma_{k}L_{ k}\right)\right]_{\gamma_{\text{lb}}}^{\gamma_{\text{ub}}},\label{eq:robbins-monro-gamma}
\end{align}where $[\gamma]_{a}^b=\min\{b, \max\{\gamma, a\}\}$ and $\{\eta_k\}$ is a set of diminishing step sizes that is selected to be:
\begin{equation}
	\eta_k=\begin{cases}
	\frac{1}{2\overline{D}_{\text{lb}}}, &k=1;\\
	\frac{1}{(k+2)\overline{D}_{\text{lb}}}, &k\geq 2. 
	\end{cases}
\end{equation}
\end{itemize}
\end{subequations}

\subsection{Theoretic Analysis}
The average AoI regret performance of the proposed policy is hard to analyze in general. As an alternative, recall that the total AoI and the sampling cost over the first $K$ frames can be computed by $\sum_{k=1}^KY_k$, where $Y_k=R_k+L_{k-1}D_k$ is the total cost in frame $k$ as pointed out by \eqref{eq:tot-frame}. The total length of the first $K$ frames is $\sum_{k=1}^KL_k$. Therefore the ratio $\overline{h}_K\triangleq\frac{\mathbb{E}\left[\sum_{k=1}^KY_k\right]}{\mathbb{E}\left[\sum_{k=1}^KL_k\right]}$ reflects the average AoI obtained by the online policy. Denote $\pi_K$ as the waiting strategy used in the $K$-th frame, i.e., $\pi_K(D)=(\gamma_K-D)^+$. We measure the performance of the proposed algorithm via the convergence rate of gap $\overline{h}_K-h_{\pi^*}$ and $h_{\pi_K}-h_{\pi^*}$, and the main results are as follow:
\begin{theorem}\label{thm:regret}
	If the transmission delay $D$ is upper bounded, i.e., $D<B<\infty$, we have:
	\begin{subequations}
	\begin{align}
    \overline{h}_K-h_{\pi^*}\leq\frac{(L_{\text{ub}}^2+C)^2}{\overline{D}\overline{D}_{\text{lb}}^2}\times\frac{1+\ln K}{K}.\label{eq:theorem4-conclusion1}
\end{align}
where $L_{\text{ub}}=B+\gamma_{\text{ub}}$. The expected difference between the expected average cost by using policy $\pi_K$ in frame $K$ and $h_{\pi^*}$ can be upper bounded by:
	\begin{align}
		\mathbb{E}\left[h_{\pi_K}-h_{\pi^*}\right]\leq\frac{(L_\text{ub}^2+C)^2}{\overline{D}\overline{D}_{\text{lb}}^2}\times\frac{1}{K}.\label{eq:theorem4-conclusion2}
	\end{align}
	\end{subequations}
\end{theorem}

Proof of Theorem~\ref{thm:regret} can be found in Appendix~\ref{pf:thm:regret}. 


\section{Simulations}
In this section we simulate three sampling policies: (1). Zero-wait policy that takes a new sample immediately when the ACK of the last sample is received, i.e.,  $\pi_{\text{zw}}(d)=0, \forall d$; (2). The optimum off-line policy $\pi^*$ with known $\mathsf{P}_D$ proposed in Section III-B; (3). The online algorithm in Section III-C. Simulations are carried out when the transmission delay $D$ follows a log-normal distribution parameterized by $\mu$ and $\sigma$, i.e., the density function 
\[f_D(d):=\frac{\mathsf{P}_D(\mathsf{d}d)}{\mathsf{d}d}=\frac{1}{d\sigma\sqrt{2\pi}}\exp\left(-\frac{(\ln d-\mu)^2}{2\sigma^2}\right). \]

\begin{figure}[t]
	\centering
	\includegraphics[width=.45\textwidth]{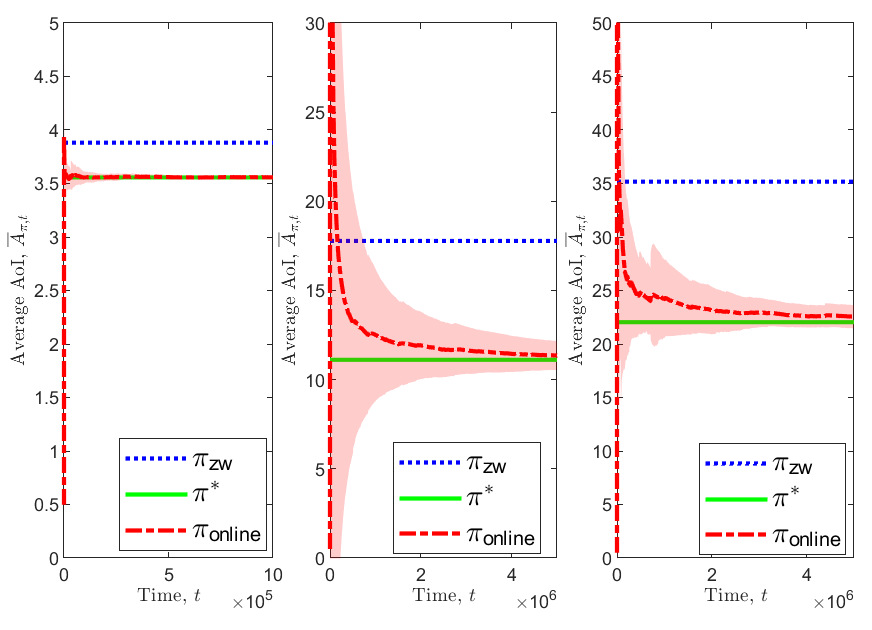}
	\caption{The average AoI $\overline{A}_{\pi, t}$ evolution with time $t$ by using various algorithm. From left to right $(\mu, \sigma)=(1, 1), (1, 1.5), (2, 1.5)$. }
	\label{fig:uncons-lognormal}
\end{figure}

First we consider $C=0$, i.e., sampling has no extra cost. Therefore, the total cost minimization problem degrades to the average AoI minimization problem considered in \cite{sun_17_tit}. The expectation $\overline{A}_{\pi, t}$ is computed by taking the  average of $\frac{1}{t}\int_{0}^tA(t)\mathsf{d}t$ over 50 simulations and the confidence region is marked in red in the figure. As is illustrated in Fig.~\ref{fig:uncons-lognormal}, the time average AoI obtained by our algorithm converges to the expected average AoI obtained by the optimum algorithm \cite{sun_17_tit} for all the parameters. Notice that a small $\sigma$ indicates the variance of the delay is small, comparing the first and second sub-plots in Fig.~\ref{fig:uncons-lognormal}, our online algorithm converges faster when the variance of transmission delay is small. 

\begin{figure}[h]
	\centering
	\includegraphics[width=.45\textwidth]{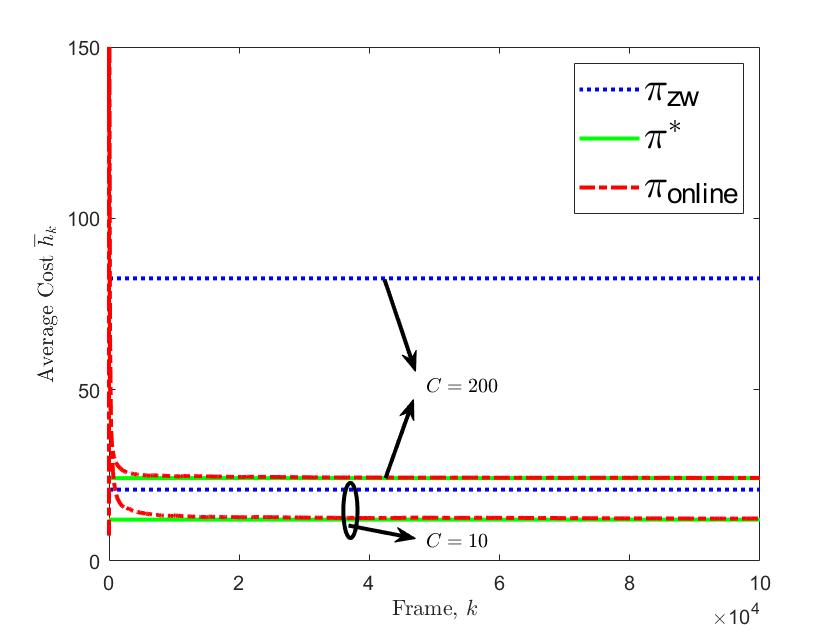}
	\caption{The average cost evolution with the number of frames $k$. The transmission delay $\mathsf{P}_D$ follows the log-normal distribution parameterized by $(\mu, \sigma)=(1, 1.5)$. }
	\label{fig:cons-lognormal}
\end{figure}

Next we study time average cost obtained by the proposed algorithm for $C>0$. Recall that $X_k$ is the cumulative AoI in frame $K$ and $C$ is the sampling cost. We plot the average cost $\overline{h}_K=\frac{\mathbb{E}\left[\sum_{k=1}^K(X_k+C)\right]}{\mathbb{E}\left[\sum_{k=1}^KL_k\right]}$ up to frame $K$ by using different algorithms in Fig.~\ref{fig:cons-lognormal}. The transmission delay $D$ follows the log-normal distribution with parameter $\mu=1$ and $\sigma=1.5$. The proposed online learning algorithm adaptively learns the optimum policy that balances the average AoI performance and sampling cost. The zero-wait policy is far from optimum when the sampling cost is large. 
\section{Conclusions}
We investigated sampling strategies to minimize the sum of average AoI and sampling cost for status update system with random transmission delay. We reformulate the problem as the optimization of a renewal-reward process and find the optimal policy when the delay distribution is known. We then propose an online sampling strategy that adaptively learns the optimum offline policy using the Robbins-Monro algorithm. It has been demonstrated that the optimality gap between the average cost of the online algorithm and the minimum cost obtained by the optimal offline policy diminishes when the number of samples goes to infinity. 

\appendices
\section{Proof of Corollary~\ref{coro:gamma-bound}}\label{pf:coro:gamma-bound}
\begin{IEEEproof}
	First we derive the lower bound of $\gamma^*$. Consider any stationary deterministic policy $\pi\in\Pi_{\text{SD}}$, we have:
	\begin{align}
	&\frac{\mathbb{E}[\frac{1}{2}(D+\pi(D))^2+C]}{\mathbb{E}[D+\pi(D)]}
	\overset{(a)}{\geq}\frac{1}{2}\frac{\mathbb{E}[D+\pi(D)]^2}{\mathbb{E}[D+\pi(D)]}\nonumber\\
	=&\frac{1}{2}\mathbb{E}[D+\pi(D)]\geq\frac{1}{2} D_{\text{lb}}=:\gamma_{\text{lb}}. 
	\end{align}
	where inequality $(a)$ is because $C\geq 0$ and the Cauchy-Schwartz inequality implies $\mathbb{E}[(D+\pi(D))^2]\geq\mathbb{E}[D+\pi(D)]^2$.
	
	To find the upper bound of $\gamma^*$, we consider zero wait policy $\pi_{\text{zw}}$ that selects $W\equiv 0, \forall D$. Since policy $\pi_{\text{zw}}$ may not be the optimum policy, we have:
	\begin{align}
	\gamma^*&\leq\frac{\mathbb{E}\left[\frac{1}{2}(D+\pi_{ \text{zw}}(D))^2+C\right]}{\mathbb{E}[D+\pi_{\text{zw}}(D)]}\nonumber\\
	&=\frac{\frac{1}{2}\mathbb{E}[D^2]+C}{\overline{D}}\leq\frac{\frac{1}{2}M_\text{ub}+C}{\overline{D}_{\text{lb}}}=:\gamma_\text{ub}.\end{align}
\end{IEEEproof}
\section{Proof of Corollary~\ref{coro:uniquesolve}}\label{pf:coro:uniquesolve}
\begin{IEEEproof}
	Recall that the mapping $\mathcal{T}(\cdot):\mathbb{R}\rightarrow\mathbb{R}$ is:
\[\mathcal{T}(\gamma):=\frac{\mathbb{E}[\frac{1}{2}(D+\tilde{\pi}_{ \gamma}^*(D))^2+C]}{\mathbb{E}[D+\tilde{\pi}_{ \gamma}^*(D)]}.\]
	and according to Theorem~\ref{thm:threshold}, the waiting time obtained by policy $\tilde{\pi}_{ \gamma}^*$ is:
	\[\tilde{\pi}_{\gamma}^*(d)=(\gamma-d)^+.\]

	Therefore, $d+\tilde{\pi}_{ \gamma}^*(d)=\max\{d, \gamma\}$, we can rewrite mapping $\mathcal{T}(\gamma)$ as follows:
	\begin{equation}
	\mathcal{T}(\gamma)=\frac{\mathbb{E}\left[\frac{1}{2}(\max\{D, \gamma\})^2+C\right]}{\mathbb{E}\left[\max\{D, \gamma\}\right]}.
	\end{equation}
	
	Since distribution $\mathsf{P}_D$ is absolutely continuous, $\mathcal{T}$ is continuous. Next, we will show that for any $\gamma\in[\gamma_{\text{lb}}, \gamma_{\text{ub}}]$, $\mathcal{T}(\gamma)$ is bounded via the following inequality:
	\begin{align}
	&\mathcal{T}(\gamma)\leq\frac{\mathbb{E}[\frac{1}{2}\left(D+\gamma\right)^2+C]}{\overline{D}}\nonumber\\
	&\leq\frac{1}{\overline{D}}\left(\frac{1}{2}M_{\text{ub}}+\overline{D}\gamma_{\text{ub}}+\gamma_{\text{ub}}^2+C\right)=:\mathcal{T}_{\text{ub}}.
	\end{align}
	
	Apparently, $\mathcal{T}_{\text{ub}}\geq\gamma_{\text{ub}}$ and the stationary point $\gamma\in[\gamma_{\text{lb}}, \mathcal{T}_{\text{ub}}]$. 
	
	We will then show mapping $\mathcal{T}(\gamma)$ has a unique stationary point $\gamma^*$. Suppose $\gamma^*$ is a stationary point, we have:
	\begin{equation}
	    \mathbb{E}\left[\frac{1}{2}(\max\{D, \gamma^*\})^2+C\right]=\gamma^*\mathbb{E}\left[\max\{D,\gamma^*\}\right].\label{eq:stationarycon}
	\end{equation}
	
The proof is divided into two cases:
	\begin{itemize}
		\item If $\gamma>\gamma^*$, we will show $\mathcal{T}(\gamma)$ is a sub-contraction mapping:
		\begin{align}
		&\mathcal{T}(\gamma)-\gamma^*=\frac{\mathbb{E}\left[\frac{1}{2}(\max\{D, \gamma\})^2+C\right]}{\mathbb{E}\left[\max\{D, \gamma\}\right]}-\gamma^*\nonumber\\
		=&\frac{\mathbb{E}\left[\frac{1}{2}(\max\{D, \gamma\})^2-\gamma\max\{D, \gamma\}+C\right]}{\mathbb{E}\left[\max\{D, \gamma\}\right]}\nonumber\\
		&+(\gamma-\gamma^*)\nonumber\\
		\overset{(a)}{<}&\frac{\mathbb{E}\left[\frac{1}{2}(\max\{D, \gamma^*\})^2-\gamma\max\{D,\gamma^*\}+C\right]}{\mathbb{E}\left[\max\{D, \gamma\}\right]}\nonumber\\
		&+(\gamma-\gamma^*)\nonumber\\
		=&\frac{\mathbb{E}\left[\frac{1}{2}(\max\{D,\gamma^*\})^2-\gamma^*\max\{D, \gamma^*\}+C\right]}{\mathbb{E}\left[\max\{D,\gamma\}\right]}\nonumber\\
		&+(\gamma^*-\gamma)\frac{\mathbb{E}\left[\max\{D,\gamma^*\}\right]}{\mathbb{E}\left[\max\{D,\gamma\}\right]}+(\gamma-\gamma^*)\nonumber\\
		\overset{(b)}{=}&(\gamma-\gamma^*)\left(1-\frac{\mathbb{E}\left[\max\{D,\gamma^*\}\right]}{\mathbb{E}\left[\max\{D,\gamma\}\right]}\right)
		\leq(\gamma-\gamma^*),\label{coro2-ub1}
		\end{align}
		where inequality (a) is because policy $\pi(D)=(\gamma-D)^+$ is the optimum policy to minimize function $g(\pi, \gamma)$; equality (b) is obtained because of \eqref{eq:stationarycon}. 
		
		We will then show $\mathcal{T}(\gamma)-\gamma^*$ is positive:
		\begin{align}
		&\mathcal{T}(\gamma)-\gamma^*\nonumber\\
		=&\frac{\mathbb{E}\left[\frac{1}{2}(\max\{D, \gamma\})^2+C\right]}{\mathbb{E}\left[\max\{D,\gamma\}\right]}-\gamma^*\nonumber\\
		=&\frac{\mathbb{E}\left[\frac{1}{2}(\max\{D,\gamma\})^2-\gamma^*\cdot\max\{D, \gamma\}+C\right]}{\mathbb{E}\left[\max\{D, \gamma\}\right]}\nonumber\\
		\overset{(c)}{\geq}&\frac{\mathbb{E}\left[\frac{1}{2}(\max\{D,\gamma^*\})^2-\gamma^*\cdot\max\{D,\gamma^*\}+C\right]}{\mathbb{E}\left[\max\{D,\gamma\}\right]}\nonumber\\
		\overset{(d)}{=}&0,\label{eq:coro2-lb1}
		\end{align}
		where inequality (c) is because policy $\tilde{\pi}_{\gamma^* }(D)=(\gamma^*-D)^+$ is the optimum policy to minimize $g(\pi, \gamma^*)$ and equality (d) is obtained because of \eqref{eq:stationarycon}. 
		
		 Since the stationary point $\gamma^*$ exists, combining \eqref{coro2-ub1} and \eqref{eq:coro2-lb1} leads to the conclusion that $\mathcal{T}(\cdot)$ is a sub-contraction mapping:
		\[|\mathcal{T}(\gamma)-\gamma^*|<(\gamma-\gamma^*), \forall \gamma>\gamma^*. \]
		
		Finally, consider that the stationary point exists and satisfies $\gamma^*<\gamma_{\text{ub}}<\infty$, the image $0<\mathcal{T}(\gamma)<\gamma_{\text{ub}}<\infty$ belongs to a compact set. According to \cite{subcontract}, we have: \begin{equation}\lim_{\tau\rightarrow\infty}\mathcal{T}^{(\tau)}(\gamma)=\gamma^*.\label{eq:positive-limit}
		\end{equation}
		
		\item If $\gamma<\gamma^*$, we will first show $\mathcal{T}(\gamma)-\gamma^*>(\gamma-\gamma^*)$:
		\begin{align}
		&\mathcal{T}(\gamma)-\gamma^*=\frac{\mathbb{E}\left[\frac{1}{2}(\max\{D, \gamma\})^2+C\right]}{\mathbb{E}\left[\max\{D,\gamma\}\right]}-\gamma^*\nonumber\\
		\overset{(e)}{>}&\frac{\mathbb{E}\left[\frac{1}{2}(\max\{D,\gamma\})^2+C\right]}{\mathbb{E}\left[\max\{D,\gamma^*\}\right]}-\gamma^*\nonumber\\
		\overset{}{=}&\frac{\mathbb{E}\left[\frac{1}{2}(\max\{D,\gamma\})^2-\gamma^*\cdot\max\{D,\gamma\}+C\right]}{\mathbb{E}\left[\max\{D,\gamma^*\}\right]}\nonumber\\
		&+\gamma^*\left(\frac{\mathbb{E}\left[\max\{D,\gamma\}\right]}{\mathbb{E}\left[\max\{D,\gamma^*\}\right]}-1\right)\nonumber\\
		\overset{(f)}{\geq}&\frac{\mathbb{E}\left[\frac{1}{2}(\max\{D,\gamma^*\})^2-\gamma^*\cdot\max\{D,\gamma^*\}+C\right]}{\mathbb{E}\left[\max\{D,\gamma^*\}\right]}\nonumber\\
		&+\gamma^*\left(\frac{\mathbb{E}\left[\max\{D,\gamma\}\right]}{\mathbb{E}\left[\max\{D,\gamma^*\}\right]}-1\right)\nonumber\\
		\overset{(g)}{=}&\frac{\gamma^*}{\mathbb{E}\left[\max\{D,\gamma^*\}\right]}\left(\mathbb{E}\left[\max\{D,\gamma\}\right]-\mathbb{E}\left[\max\{D,\gamma^*\}\right]\right)\nonumber\\
		\overset{(h)}{\geq}&(\gamma-\gamma^*),
		\label{eq:nega-limit}
		\end{align}
		where inequality (e) is because $\gamma<\gamma^*$ implies $\mathbb{E}\left[\max\{D,\gamma\}\right]<\mathbb{E}\left[\max\{D,\gamma^*\}\right]$; inequality (f) is because policy $\pi(D)=(\gamma^*-D)^+$ is the optimum policy to minimize function $g(\pi, \gamma^*)$; equality (g) is obtained because of \eqref{eq:stationarycon}; inequality (h) is obtained because $ \frac{\gamma^*}{\mathbb{E}\left[\max\{D, \gamma^*\}\right]}< 1$ and $\gamma<\gamma^*$ implies \begin{align}&\mathbb{E}\left[\max\{D,\gamma\}\right]-\mathbb{E}\left[\max\{D,\gamma^*\}\right]\nonumber\\
		\geq&\mathbb{E}\left[(\gamma-\gamma^*)\}\right]=\gamma-\gamma^*.\nonumber
		\end{align}
		
		For the case that $\gamma<\gamma^*$, denote $\iota$ be the stopping time that: \begin{equation}
		\iota(\gamma)=\arg\min_{\tau\in\mathbb{N}^+}\{\mathcal{T}^{(\tau)}(\gamma)\geq\gamma^*\}.\nonumber\end{equation} 
		
		If $\iota(\gamma)<\infty$, then $\mathcal{T}^{(\iota(\tau))}>\gamma^*$, and according to \eqref{eq:positive-limit} we have \[\lim_{\tau\rightarrow\infty}\mathcal{T}^{(\tau)}(\mathcal{T}^{(\iota(\gamma))}(\gamma))=\gamma^*.\] 
		Otherwise, if $\iota(\gamma)=\infty$, due to \eqref{eq:nega-limit}, the mapping is subcontract. Since the stationary point exists, $\lim_{\tau\rightarrow\infty}\mathcal{T}^{(\tau)}(\gamma)=\gamma^*$. 
	\end{itemize}	
	
\end{IEEEproof}

\section{Proof of Theorem~\ref{thm:regret}}\label{pf:thm:regret}
\begin{IEEEproof}
    First, recall that the ratio $\gamma_k$ used in any frame $k$ is upper bounded by $\gamma_{\text{ub}}$, since the transmission delay is bounded $D_k\leq B$, the frame length $L_k$ and the reward $R_k$ can be upper bounded by:
    \begin{subequations}
    \begin{align}
    	L_k&\leq D_k+(\gamma-D_k)^+\leq B+\gamma_{\text{ub}}=:L_{\text{ub}},\label{lub}\\
    	R_k&=\frac{1}{2}L_k^2+C\leq L_{\text{ub}}^2+C.\label{rub}
    \end{align}
    \end{subequations}
    
    Denote $\overline{L}^*:=\mathbb{E}[D+\pi^*(D)]$ and $\overline{R}^*:=\mathbb{E}[\frac{1}{2}(D+\pi^*(D))^2+C]$ to be the expected frame length and the reward in each frame using the optimum policy $\pi^*$. To proceed, we provide Lemma~\ref{lemma:cons-1} and Lemma~\ref{lemma:cons-2new}, the proofs are provided in Appendix~\ref{pf:lemma:cons-1} and \ref{pf:lemma:cons-2new}, respectively. 
\begin{lemma}\label{lemma:cons-1}
	The expected frame length $\mathbb{E}[L_{k}|\mathcal{F}_{k-1}]$ and the expected reward $\mathbb{E}[R_{k}|\mathcal{F}_{k-1}]$ of frame $k$ satisfies:
	\begin{subequations}
		\begin{align}
		&\mathbb{E}\left[R_k-\gamma_k L_{k}|\mathcal{F}_{ k-1}\right]\leq(\gamma^{ *}-\gamma_{ k})\overline{L}^{*},\label{eq:lemmacons-1-eqn1}\\
		&\mathbb{E}\left[R_k-\gamma^{ *}L_{k}|\mathcal{F}_{ k-1}\right]\leq\!-\!(\gamma^{*}-\gamma_{k})\left(\mathbb{E}[L_{ k}|\mathcal{F}_{ k-1}]-\overline{L}^{*}\right)\!. \label{eq:lemmacons-1-eqn2}
		\end{align}
	\end{subequations}
\end{lemma}

\begin{lemma}{\label{lemma:cons-2new}}
    Recall that $Y_k=R_k+L_{k-1}D_k$ is the sum of cumulative AoI and extra cost within frame $k$ according to \eqref{eq:tot-frame}.  Denote \[\Delta_K:= \mathbb{E}\left[\sum_{k=1}^{K}(Y_k-\gamma^{*}L_{k})\right].\] 
    
    Then, $\Delta_K$ can be upper bounded by:
	\begin{align}
	\Delta_K\leq\mathbb{E}\left[\sum_{k=1}^K(\gamma^{*}-\gamma_{ k})^2\right].\label{eq:lemma-2}
	\end{align}
\end{lemma}

Notice that the average cost deviation $\overline{h}_K-h_{\pi^*}=\frac{1}{\mathbb{E}[\sum_{k=1}^KL_k]}\Delta_K$, it is suffice to study the convergence behavior of $\mathbb{E}\left[\sum_{k=1}^K(\gamma^{*}-\gamma_{ k})^2\right]$. The next lemma bound the expected difference between $\gamma_k$ and $\gamma^*$ in frame $k$, whose proof is provided in Appendix~\ref{pf:lemma:gammak-cons}:
\begin{lemma}\label{lemma:gammak-cons}
    The expected difference between $\gamma_k$ and $\gamma^*$ can be upper bounded by:
    \begin{equation}\mathbb{E}[(\gamma_{ k}-\gamma^{ *})^2]\leq\frac{1}{k}\frac{(L_{\text{ub}}^2+C)^2}{\overline{D}_{\text{lb}}^2}.\label{eq:gamma-diminish}
    \end{equation}
\end{lemma}

With the above lemmas, we can verify the two conclusions in Theorem~\ref{thm:regret} respectively:

\textit{Proof of \eqref{eq:theorem4-conclusion1}}: Plugging \eqref{lemma:gammak-cons} into inequality \eqref{eq:lemma-2} from Lemma~\ref{lemma:cons-2new}, we can first upper bound $\Delta_K$ by:
\begin{align}
\Delta_K&\leq\frac{(L_{\text{ub}}^2+C)^2}{\overline{D}_{\text{lb}}^2}\left(\sum_{k=1}^K\frac{1}{k}\right)\nonumber\\
&\overset{(a)}{\leq}\frac{(L_{\text{ub}}^2+C)^2}{\overline{D}_{\text{lb}}^2}\left( 1+\int_{k=1}^{K}\frac{1}{k}\mathsf{d}k\right)\nonumber\\
&=\frac{(L_{\text{ub}}^2+C)^2}{\overline{D}_{\text{lb}}^2}\left(1+\ln K\right),
\end{align}
where inequality $(a)$ is because $\frac{1}{k}\leq\int_{k-1}^{k}\frac{1}{x}\mathsf{d}x$. 

Notice that:
\begin{align}
\overline{h}_K-h_{\pi^*}=&\frac{1}{\mathbb{E}\left[\sum_{k=1}^KL_k\right]}\Delta_K\nonumber\\
\overset{(b)}{\leq}&\frac{(L_{\text{ub}}^2+C)^2}{\overline{D}\overline{D}_{\text{lb}}^2}\frac{\left(1+\ln K\right)}{K},
\end{align}
where inequality $(b)$ is because  \[\mathbb{E}\left[\sum_{k=1}^KL_k\right]\geq \mathbb{E}\left[\sum_{k=1}^KD_k\right]\geq K\overline{D}_{\text{lb}}.\]

\textit{Proof of \eqref{eq:theorem4-conclusion2}}: Recall that the average cost using policy $\pi_K$ can be computed by \[h_{\pi_K}=\frac{\mathbb{E}[\frac{1}{2}((\gamma_K-D)^++D)^2+C]}{\mathbb{E}[(\gamma_K-D)^++D]}+\overline{D},\] which is smaller or equal to the minimum cost, i.e., $h_{\pi_K}\geq h_{\pi^*}$. Therefore for each $\gamma_K$, we can upper bound the gap between policy $\pi_K$ and $\pi^*$ by:
\begin{align}
	&h_{\pi_K}-h_{\pi^*}=h_{\pi_K}-h_{\pi^*}\nonumber\\
	=&\frac{\mathbb{E}\left[\frac{1}{2}((\gamma_K-D)^++D)^2+C\right]}{\mathbb{E}[(\gamma_K-D)^++D]}-\gamma^*\nonumber\\
	=&\frac{\mathbb{E}\left[\frac{1}{2}((\gamma_K-D)^+\!+\!D)^2\!+\!C\!-\!\gamma_K((\gamma_K-D)^++D)\right]}{\mathbb{E}\left[(\gamma_K-D)^++D\right]}\nonumber\\
	&+(\gamma_K-\gamma^*)\nonumber\\
	\leq&\frac{\mathbb{E}[\frac{1}{2}((\gamma^*-D)^++D)^2+C-\gamma_K((\gamma^*-D)^++D)]}{\mathbb{E}[(\gamma_K-D)^++D]}\nonumber\\
	&+(\gamma_K-\gamma^*)\nonumber\\
	\leq&\frac{(\gamma^*-\gamma_K)\mathbb{E}[(\gamma_K-D)^++D]}{\mathbb{E}[(\gamma_K-D)^++D]}+(\gamma_K-\gamma^*)\nonumber\\
	=&\frac{(\gamma^*-\gamma_K)}{\mathbb{E}[(\gamma_K-D)^++D]}\mathbb{E}[(\gamma_K-D)^+-(\gamma^*-D)^+]\nonumber\\
	\leq&\frac{1}{\overline{D}}(\gamma_K-\gamma^*)^2.
\end{align}

Plugging \eqref{eq:gamma-diminish} from Lemma~\ref{lemma:gammak-cons} into the above equation, we have \[\mathbb{E}[h_{\pi_K}-h_{\pi^*}]\leq\frac{(L_\text{ub}^2+C)^2}{\overline{D}\overline{D}_{\text{lb}}^2}\frac{1}{K}. \]

And this completes the proof of Theorem~\ref{thm:regret}. 
\end{IEEEproof}

\section{Proof of Lemma~\ref{lemma:cons-1}}\label{pf:lemma:cons-1}
\begin{IEEEproof}
	Notice that in each frame $k$, the waiting time $W_k$ is chosen to minimize the objective function \eqref{eq:consopt}, therefore we have:
	\begin{align}
	&\mathbb{E}\left[R_k-\gamma_{ k}L_{k}|\mathcal{F}_{ k-1}\right]
	\overset{(a)}{\leq}(\overline{R}^*-\gamma_k\overline{L}^{*})\nonumber\\
	\overset{}{=}&(\overline{R}^*-\gamma^*\overline{L}^{*})+(\gamma^{*}-\gamma_k)\overline{L}^*
	\overset{(b)}{=}(\gamma^{*}-\gamma_k)\overline{L}^*,\label{eq:lemm7-prev}
	\end{align}
	where equality $(a)$ is because policy $\pi_{k}$ used in frame $k$ satisfies $g(\pi_{k}, \gamma_k)\leq g(\pi^{*}, \gamma_k)$. Equality $(b)$ is obtained because on the stationary point $\gamma^*$ we have $\overline{R}^{*}=\gamma^{ *}\overline{L}^{*}$. This verifies the first inequality in Lemma~\ref{lemma:cons-1}.
	
	Then, adding $(\gamma_{ k}-\gamma^{*})\mathbb{E}[L_{ k}|\mathcal{F}_{k-1}]$ to both sides of \eqref{eq:lemm7-prev}, we have:
	\begin{align}
	&\mathbb{E}\left[R_k-\gamma^{*}L_{ k}|\mathcal{F}_{k-1}\right]\leq(\gamma_{ k}-\gamma^{*})\mathbb{E}\left[L_{ k}-\overline{L}^{*}|\mathcal{F}_{k-1}\right].
	\end{align}
	which verifies the second inequality in Lemma \ref{lemma:cons-1}. \end{IEEEproof}

\section{Proof of Lemma~\ref{lemma:cons-2new}}\label{pf:lemma:cons-2new}
\begin{IEEEproof}
%
%
To find the upper bound of $\Delta_k$, first we add $\mathbb{E}[L_{k-1}D_{ k}|\mathcal{F}_{k-1}]$ on both sides of \eqref{eq:lemmacons-1-eqn2}. By replacing $R_k+L_{k-1}D_k$ with $Y_k$, we then have the following inequality:
    \begin{align}
        &\mathbb{E}[Y_{k}-\gamma^{*}L_{ k}|\mathcal{F}_{k-1}]\nonumber\\
        \leq&-(\gamma^{ *}-\gamma_{ k})\left(\mathbb{E}[L_{ k}|\mathcal{F}_{ k-1}]-\overline{L}^{ *}\right)+\mathbb{E}[L_{ k-1}|\mathcal{F}_{ k-1}]\overline{D}\nonumber\\
        \overset{(d)}{\leq}&(\gamma^{*}-\gamma_{ k})^2+\mathbb{E}[L_{ k-1}|\mathcal{F}_{k-1}]\overline{D},\label{eq:lemm4-2}
    \end{align}
    where inequality $(d)$ is because \begin{align}\mathbb{E}[L_k-\overline{L}^*|\mathcal{F}_{k-1}]&=\mathbb{E}\left[(\gamma_k-D)^+-(\gamma^*-D)^+\right]\nonumber\\
    &\leq|\gamma_k-\gamma^*|. 
    \end{align}
    
    Summing up inequality \eqref{eq:lemm4-2} from frame $k=1$ to $K$ and taking the expectation with respect to $\mathcal{F}_{K-1}$, we have:
    \begin{align}
    &\mathbb{E}\left[\sum_{k=1}^K\left(Y_{k}-\left(\gamma^{*}+\overline{D}\right)L_{k}\right)\right]\nonumber\\
    \leq&\mathbb{E}\left[\sum_{k=1}^K(\gamma^{*}-\gamma_{ k})^2\right]-\mathbb{E}[L_{K}]\overline{D}. 
    \end{align}
    
    And this completes the proof of Lemma~\ref{lemma:cons-2new}. 
\end{IEEEproof}
\section{Proof of Lemma~\ref{lemma:gammak-cons}}\label{pf:lemma:gammak-cons}
\begin{IEEEproof}
For simplicity, denote \begin{align}
z_{k+1}:=\gamma_{k}+\eta_k(R_k-\gamma_k L_{ k}). 
\end{align}

Since $\gamma_{k+1}=[z_{ k+1}]_{\gamma_{\text{lb}}}^{\gamma_{\text{ub}}}$ and the optimum ratio $\gamma^*\in[\gamma_{\text{lb}}, \gamma_{\text{ub}}]$, we can bound the derivation $(\gamma_{k+1}-\gamma^*)^2$ using $(z_{k+1}-\gamma^*)^2$ through the following inequality:
\begin{equation}
(\gamma_{k+1}-\gamma^{*})^2=([z_{ k+1}]_{\gamma_{\text{lb}}}^{\gamma_{\text{ub}}}-[\gamma^{*}]_{\gamma_{\text{lb}}}^{\gamma_{\text{ub}}})\leq(z_{k+1}-\gamma^{*})^2. \label{eq:lemm4-gammtoz}
\end{equation}

Next, recall the update rule given in \eqref{eq:robbins-monro-gamma}, we have:
\begin{align}
&\frac{1}{2}(z_{k+1}-\gamma^{ *})^2\nonumber\\
=&\frac{1}{2}\left(\gamma_{k}-\gamma^{ *}+\eta_k\left(R_k-\gamma_{ k}L_{ k}\right)\right)^2\nonumber\\
=&\frac{1}{2}(\gamma_{k}-\gamma^{ *})^2+\frac{1}{2}\eta_{k}^2\left(R_k-\gamma_{k} L_{k}\right)^2\nonumber\\
&+\eta_k(\gamma_{k}-\gamma^{ *})\left(R_k-\gamma_{k}L_{k}\right)\nonumber\\
\leq&\frac{1}{2}(\gamma_{k}-\gamma^{ *})^2+\frac{1}{2}\eta_k^2(L_{\text{ub}}^2+C)^2\nonumber\\
&+\eta_k(\gamma_{k}-\gamma^{ *})\left(R_k-\gamma_{k} L_{k}\right),
\label{eq:step-dminimish1}
\end{align}
where the last inequality is obtained because $R_k=\frac{1}{2}L_k^2+C\leq L_{\text{ub}}^2$ and $\gamma_kL_k\leq L_{\text{ub}}^2$. Then, conditioned on filtration $\mathcal{F}_{k-1}$ and take the expectation on both sides of~\eqref{eq:step-dminimish1}, we have:
\begin{align}
&\frac{1}{2}\mathbb{E}\left[(z_{ k+1}-\gamma^{ *})^2|\mathcal{F}_{k-1}\right]\nonumber\\
\leq &\frac{1}{2}(\gamma_{k}-\gamma^{ *})^2+\frac{1}{2}\eta_k^2(L_{\text{ub}}^2+C)^2\nonumber\\
&+\eta_k(\gamma_{k}-\gamma^{ *})\mathbb{E}\left[R_k-\gamma_{k}L_{ k}|\mathcal{F}_{ k-1}\right].\label{eq:lemm4-drift}
\end{align}

We then proceed to bound the last term in inequality~\eqref{eq:lemm4-drift}. The analysis is divided into two cases:
\begin{itemize}
	\item If the current $\gamma_{ k}-\gamma^{*}\geq 0$, by plugging \eqref{eq:lemmacons-1-eqn1} into the above equation, we have:
	\begin{align}
	&(\gamma_{k}-\gamma^{ *})\mathbb{E}[R_k-\gamma_{k}L_{ k}|\mathcal{F}_{k-1}]\nonumber\\
	\leq&-(\gamma_{k}-\gamma^{ *})^2\overline{L}^{ *}\leq-(\gamma_k-\gamma^*)^2\overline{D},
	\label{eq:lemm4-ub3-1}
	\end{align}
	where the last inequality is obtained because $\overline{L}^*\geq\overline{D}$. 
	
	\item  If the current $\gamma_{ k}-\gamma^{*}\leq 0$, we can upper the last term in inequality \eqref{eq:lemm4-drift} as follows:
	\begin{align}
	&(\gamma_{k}-\gamma^{ *})\mathbb{E}[R_k-\gamma_{k}L_{ k}|\mathcal{F}_{ k-1}]\nonumber\\
	=&(\gamma_{k}-\gamma^{ *})\mathbb{E}[R_k-\gamma^{ *}L_{ k}|\mathcal{F}_{k-1}]\nonumber\\
	&-(\gamma_{k}-\gamma^{ *})^2\mathbb{E}[L_{ k}|\mathcal{F}_{k-1}]\nonumber\\
	\overset{(a)}{\leq}&(\gamma_{k}-\gamma^{ *})(\overline{R}^{ *}-\gamma^{*}\overline{L}^{*})-(\gamma_{k}-\gamma^{*})^2\mathbb{E}[L_{ k}|\mathcal{F}_{k-1}]\nonumber\\
	=&-(\gamma_{k}-\gamma^{ *})^2\mathbb{E}[L_{ k}|\mathcal{F}_{k-1}]\nonumber\\
	\overset{(b)}{\leq}&-(\gamma_{k}-\gamma^{ *})^2\overline{D},\label{eq:lemm4-ub3-2}
	\end{align}
	where inequality $(a)$ is because $\mathbb{E}[R_k-\gamma^*L_k|\mathcal{F}_{k-1}]\geq\overline{R}^*-\gamma^*\overline{L}^*=0$ and inequality $(b)$ is because $\mathbb{E}[L_k|\mathcal{F}_{k-1}]\geq\overline{D}$.  
\end{itemize}

Plugging \eqref{eq:lemm4-ub3-1} and \eqref{eq:lemm4-ub3-2} into \eqref{eq:lemm4-drift} yields:
\begin{align}
&\frac{1}{2}\mathbb{E}\left[(z_{k+1}-\gamma^{*})^2|\mathcal{F}_{k-1}\right]\nonumber\\
=&\left(\frac{1}{2}-\eta_k\overline{D}\right)(\gamma_k-\gamma^*)^2+\frac{1}{2}\eta_k^2(L_{\text{ub}}^2+C)^2\nonumber\\
\leq &\left(\frac{1}{2}-\eta_k\overline{D}_{\text{lb}}\right)(\gamma_k-\gamma^*)^2+\frac{1}{2}\eta_k^2(L_{\text{ub}}^2+C)^2.\label{eq:condeq}
\end{align}

By taking the expectation with respect to filtration $\mathcal{F}_{k-1}$ on both sides of inequality \eqref{eq:condeq} and then plugging it into \eqref{eq:lemm4-gammtoz}, we can upper bound $\mathbb{E}[(\gamma_{k+1}-\gamma^*)^2]$ through:
\begin{align}
&\frac{1}{2}\mathbb{E}\left[(\gamma_{k+1}-\gamma^{*})^2\right]\overset{(c)}{\leq}\frac{1}{2}\mathbb{E}[(z_{k+1}-\gamma^*)^2]\nonumber\\
\leq&\frac{1}{2}\left(1-2\eta_k\overline{D}_{\text{lb}}\right)\mathbb{E}\left[(\gamma_{ k}-\gamma^{ *})^2\right]+\frac{1}{2}\eta_k^2(L_{\text{ub}}^2+C)^2,
\end{align}
where inequality $(c)$ is obtained because of inequality \eqref{eq:lemm4-gammtoz}. 

Recall that the stepsizes are selected through $\eta_1=\frac{1}{2\overline{D}_{\text{lb}}}$ and $\eta_k=\frac{1}{(k+2)\overline{D}_{\text{lb}}}, \forall k>1$, we can then show by induction that 
\begin{equation}
\frac{1}{2}\mathbb{E}[(\gamma_{k}-\gamma^{ *})^2]\leq\frac{1}{2k}\frac{(L_{\text{ub}}^2+C)^2}{\overline{D}_{\text{lb}}^2}. \label{eq:theorm4-step-diminish}
\end{equation}

Detailed proofs for \eqref{eq:theorm4-step-diminish} are as follow:

\begin{itemize}
	\item When $k=2$, by choosing $\eta_1=\frac{1}{2\overline{D}_{\text{lb}}}$ we have
	\[\frac{1}{2}\mathbb{E}[(\gamma_2-\gamma^*)^2]\leq\frac{1}{8}\frac{(L_{\text{ub}}^2+C)^2}{\overline{D}_{\text{lb}}^2}\leq\frac{1}{4}\frac{(L_{\text{ub}}^2+C)^2}{\overline{D}_{\text{lb}}^2}.\]
	
	\item When $k>2$, assuming that $\frac{1}{2}\mathbb{E}[(\gamma_{k}-\gamma^*)^2]\leq\frac{1}{2k}\frac{(L_{\text{ub}}^2+C)^2}{\overline{D}_{\text{lb}}^2}$, recall that the stepsize $\eta_k$ is chosen to be $\eta_k=\frac{1}{(k+2)\overline{D}_{\text{lb}}}$, we have:
	\begin{align}
	\hspace{-0.5cm}&\frac{1}{2}\mathbb{E}\left[(\gamma_{k+1}-\gamma^*)^2\right]\nonumber\\
	\hspace{-0.5cm}\leq&\left(\frac{1}{2}-\eta_k\overline{D}_{\text{lb}}\right)(\gamma_k-\gamma^*)^2+\frac{1}{2}\eta_k^2(L_{\text{ub}}^2+C)^2\nonumber\\
	\hspace{-0.5cm}\leq&\left(1-\frac{2}{k+2}\right)\frac{1}{2k}\frac{(L_{\text{ub}}^2+C)^2}{\overline{D}_{\text{lb}}^2}+\frac{1}{2}\frac{1}{(k+2)^2}\frac{(L_{\text{ub}}^2+C)^2}{\overline{D}_{\text{lb}}^2}\nonumber\\
	\hspace{-0.5cm}=&\frac{1}{2}\left(\frac{1}{k+2}+\frac{1}{(k+2)^2}\right)\frac{(L_{\text{ub}}^2+C)^2}{\overline{D}_{\text{lb}}^2}\nonumber\\
	\hspace{-0.5cm}=&\frac{1}{2}\frac{k+3}{(k+2)^2}\frac{(L_{\text{ub}}^2+C)^2}{\overline{D}_{\text{lb}}^2}\nonumber\\
	\hspace{-0.5cm}\overset{}{\leq}&\frac{1}{2}\frac{1}{(k+1)}\frac{(L_{\text{ub}}^2+C)^2}{\overline{D}_{\text{lb}}^2},
	\end{align}
	where the final inequality is obtained because $(k+1)(k+3)\leq (k+2)^2$. 
\end{itemize}
\end{IEEEproof}

\bibliography{bibfile}
\end{document}